\documentclass{article}
\usepackage{graphicx}
\usepackage{authblk}
\usepackage[section]{placeins}

\begin{document}

%\title{Interplay between mutation rate and genome size to keep mutational load in check drives genome size evolution in finite asexual populations}
\title{Evolution of genome size in asexual populations}

\author[1,2]{Aditi Gupta\thanks{aditi9783@gmail.com}}
\author[1,2]{Thomas LaBar}
\author[3]{Michael Miyagi}
\author[1,2,4]{Christoph Adami}
\affil[1]{BEACON Center for the Study of Evolution in Action, Michigan State University, East Lansing, MI 48824}
\affil[2]{Department of Microbiology and Molecular Genetics, Michigan State University, East Lansing, MI 48824}
\affil[3]{Department of Integrative Biology, the University of Texas at Austin, Austin, TX}
\affil[4]{Department of Physics and Astronomy, Michigan State University, East Lansing, MI 48824}
%\author{Aditi Gupta\affil{1}{BEACON Center for the Study of Evolution in Action, Michigan State University, East Lansing, MI 48824}\affil{2}{Department of Microbiology and Molecular Genetics, Michigan State University, East Lansing, MI 48824},
%Thomas LaBar\affil{1}{}\affil{2}{}\affil{3}{Program in Ecology, Evolutionary Biology, and Behavior}
%Michael Miyagi\affil{4}{Department of Integrative Biology, the University of Texas at Austin, Austin, TX}
%\and
%Christoph Adami\affil{1}{}\affil{2}{}\affil{5}{Department of Physics and Astronomy, Michigan State University, East Lansing, MI 48824}}
\date{}

\maketitle

%%%Newly updated.
%%% If significance statement need, then can use the below command otherwise just delete it.
%\significancetext{Understanding genome evolution is important because it explains how early living organisms evolved into the complex and diverse set we see today. By performing digital evolution experiments that mimic Darwinian dynamics, we show that genome expansion via beneficial insertions is a prerequisite for evolving phenotypic complexity. However, since most mutations are deleterious, high mutation rate increases the mutational load of an evolving population, forcing a reduction in genome size. Since majority of investigations on genome size evolution depend on empirical data from eukaryotes that have complex genome-editing mechanisms, our analyses highlight the roles that universal processes such as indels and base substitutions play in genome evolution.} % limit: 120, current: 107 words

%\begin{article}
\begin{abstract} % limit: 250
{Genome sizes have evolved to vary widely, from 250 bases in viroids to 670 billion bases in amoeba. This remarkable variation in genome size is the outcome of complex interactions between various evolutionary factors such as point mutation rate, population size, insertions and deletions, and genome editing mechanisms that may be specific to certain taxonomic lineages. While comparative genomics analyses have uncovered some of the relationships between these diverse evolutionary factors, we still do not understand what drives genome size evolution. Specifically, it is not clear how primordial mutational processes of base substitutions, insertions, and deletions influence genome size evolution in asexual organisms. Here, we use digital evolution to investigate genome size evolution by tracking genome edits and their fitness effects in real time. In agreement with empirical data, we find that mutation rate is inversely correlated with genome size in asexual populations. We show that at low point mutation rate, insertions are significantly more beneficial than deletions, driving genome expansion and acquisition of phenotypic complexity. Conversely, high mutational load experienced at high mutation rates inhibits genome growth, forcing the genomes to compress genetic information. Our analyses suggest that the inverse relationship between mutation rate and genome size is a result of the tradeoff between evolving phenotypic innovation and limiting the mutational load.}
%We evolved asexual populations at six different point mutation rates (but fixed indel rate and population size) in digital life platform Avida and found agreement with the empirical observation that the genome size is negatively correlated with point mutation rate. We further showed that genome expansion was largely achieved by beneficial insertions and was a prerequisite for evolving phenotypic innovation. However, at high mutation rates, the selection pressure to reduce mutational load caused genomes to shrink. This reduction in genome size was achieved by compressing the genetic information onto few genomic sites, which exacerbated the fitness cost of deleterious mutation, further compounding the mutational load and increasing the selection pressure to keep genomes smaller at the expense of phenotypic complexity. Although this study focused on asexual populations, digital life simulations can be used to study genome size evolution in eukaryotes by including retrotransposition and recombination in the artificial life platform.}
\end{abstract}

%\keywords{genome size | evolution | mutation rate | Avida}

%\abbreviations{TE: transposable element}
\section{Introduction}

Genome sizes evolve by various mechanisms, some of which are common to all domains of life (insertions and deletions) while others are seen in some taxonomic groups more than others (horizontal gene transfer in bacteria and transposable element activity in eukaryotes). While one might think that genome expansion leads to the acquisition of more protein-coding genes and functions, genome size does not strongly correlate with organismal complexity (the C-value paradox). Whole-genome sequencing data provide some explanation for this paradox: appreciable variation in eukaryotic genome sizes has been attributed to ploidy~\cite{Taftetal2007}, and to expansion of non-coding DNA such as introns, intergenic regions, and repeats~\cite{LynchConery2003}. Yet, genome size also positively correlates with the number of protein-coding genes~\cite{LynchConery2003}, suggesting that larger genome size \textit{is} a prerequisite for gaining new genes that could lead to phenotypic innovation.

%Simultaneous action of various determinants of genome size confound establishing their individual roles in genome evolution. Yet, some of these determinants preceded others as evolution gave rise to taxa with increasingly complex mechanisms of genome editing such as transposable element (TE) activity. 
Mutation rate, insertions and deletions (indels), and population size are three factors seen across the tree of life that are thought to influence genome size evolution. The negative correlation between genome size and point mutation rate is observed across the tree of life, from viruses to \textit{Homo sapiens}~\cite{Sniegowskietal2000}. However, a recent analysis based on more taxa found that this inverse relationship holds true only for prokaryotes and viruses, and that genome size and mutation rate are instead positively correlated in eukaryotes~\cite{Lynch2010}. 
%The genomic mutation rate itself is under selection to optimize the rate of adaptation~\cite{Sniegowskietal2000, MetzgarWills2000, Giraudetal2001}, underlining the direct influence of point mutation rate on genome size. 
High point mutation rate forces viruses to maintain small genome sizes in an effort to limit the number of deleterious mutations~\cite{Holmes2003}. This selection pressure to reduce genome size is so strong that viruses eliminate non-functional sequences inserted into their genomes~\cite{Zwartetal2014} and lose an essential gene if it is transferred to the host genome~\cite{Tromasetal2014}. This suggests that the point mutation rate and the evolution of genome size are inherently intertwined.

Population size, together with the point mutation rate and genome size, determines the mutation supply rate in an evolving population: if too many mutations are occurring, then reduction in any or all of point mutation rate, genome size, and population size can lower the mutation supply rate. Moreover, the effect of genetic drift is enhanced and purifying selection is weakened in small populations, allowing non-beneficial genome edits to persist for generations~\cite{Kuoetal2009}. Lynch and Conery postulate that these---initially nonadaptive---edits can become a source of phenotypic innovation later on~\cite{LynchConery2003}. In symbiotic bacteria, small population size and asexual reproduction cause bacterial genomes to shrink to an extent that they are 2-4 times smaller than the smallest genome seen in an independent-living organism~\cite{McCutcheonMoran2012}. In contrast, large population sizes in microbial populations weaken the effect of random drift, preventing accumulation of non-functional DNA and genome growth~\cite{Lynch2006}.

In addition to point mutation rate and population size, biases in patterns of insertions and deletions (indel spectra) have been suspected to contribute to the variation in genome sizes we see today~\cite{Vinogradov2004}. DNA loss via deletions is purported to be important in determining genome size, but this perspective is derived from analysis of a small number of eukaryotic genomes~\cite{Petrovetal2000,Gregory2004}. Strong deletion bias was found in 12 bacterial species as well~\cite{KuoOchman2009}, the majority of which have transposable element (TE) activity. Thus, it is likely that deletions outnumber insertions in taxa where TE proliferation leads to significant increases in non-functional DNA. This explanation, however, does not apply to genome size evolution in early living organisms and in taxa where TE activity is absent, and it is not clear how primordial genome editing mechanisms shaped the diversity in genome sizes we see today.

%Since the factors influencing genome size evolution in asexual populations are either complex or unknown,
Digital evolution provides an apt platform for understanding the evolutionary processes that determine genome size. While naturally evolving biological systems can take a very long time to show observable changes, short generational time of digital organisms significantly reduces the time-scale of experiments to study evolutionary processes~\cite{OfriaWilke2004, Adamietal2000, Batutetal2013}. In the Avida artificial life platform, these digital organisms are simple computer programs that compete for resources to replicate via a mutation prone process (see Methods and Supplementary text), thus evolving under Darwinian dynamics~\cite{OfriaWilke2004, Adamietal2000, Lenskietal2003}. 
%Another artificial life system, Aevol, is based on genetic algorithms and agent based modeling wherein the digital organisms evolve under Darwinian dynamics~\cite{Batutetal2013}. 
The ability to control the mutation rate, genome sizes (length of the program), and population size allows inquiry into the impact of mutation rate and indel spectra on evolution of genome size. Avida has been previously used to test many evolutionary hypotheses that are difficult to test via biological experimental evolution, such as the evolution of genomic complexity~\cite{Adamietal2000}, `survival of the flattest' effect in genotypes evolving at high mutation rates ~\cite{Wilkeetal2001}, co-evolution as a driving force for higher phenotypic complexity and evolvability~\cite{Zamanetal2014}, the time-dependent effect of genetic robustness on evolvability~\cite{Elenaetal2008}, and how standing genetic variation and environment influence evolutionary response to an environmental stimuli~\cite{ODonnelletal2014}. We used Avida to investigate genome size evolution because in addition to tracking genome edits and their fitness effects, it records evolution of phenotypic traits and thus can be used to interpret consequences of genome size evolution on phenotypic complexity.

Because avidians reproduce asexually and lack mechanisms of genome expansion such as TE activity, their evolutionary dynamics is most similar to that of viruses and prokaryotes. Thus, to examine the mechanisms of genome size evolution in asexual populations, we evolved populations of avidians at a range of mutation rates and followed the changes in their genome lengths, population fitness, genetic information, and phenotypic outcomes. Our results confirm that the genome size is negatively correlated with mutation rate. By tracking the changes in the genome size and the fitness effects of insertions and deletions that cause these changes, we find that insertions drive genome growth at low mutation rates, contributing to the evolution of phenotypic complexity via a two-step process: genome expansion followed by repurposing of the extra DNA to evolve new traits. Finally, we show that mutational load due to high mutation rate increases the selection pressure for reducing the genome size, resulting in smaller genomes with high information density. We conclude that genome size evolution is the result of a compromise between acquiring phenotypic complexity and restricting the mutational load.

%Due to absence of sexual reproduction and recombination in our experimental setting, this study highlights the role of base substitutions and indel spectra on genome evolution in asexual populations.

\section{Results and Discussion}

\subsection{Mutation rate is negatively correlated with genome size}

\begin{figure}[!h]
%    \centering
    \includegraphics[width=1.0\textwidth]{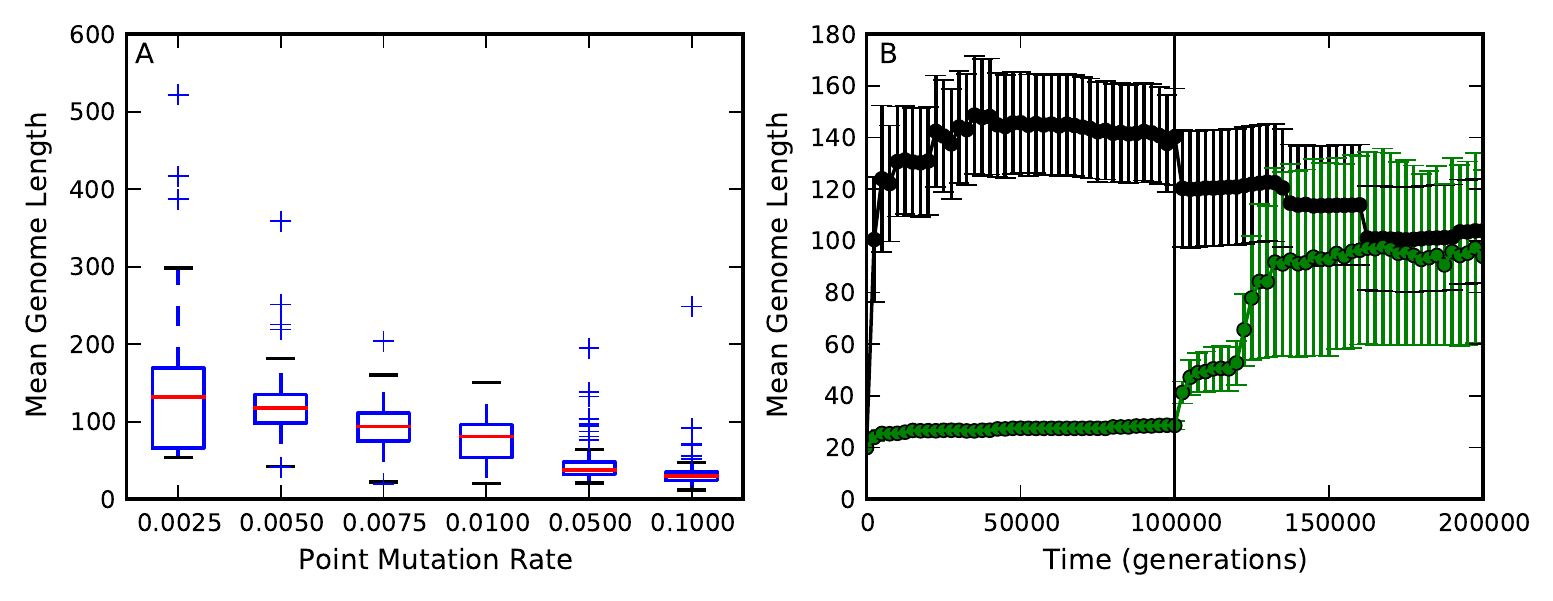}
    \caption{Point mutation rate is a strong determinant of genome size. A: Genome size and mutation rate are negatively correlated in asexual populations. The initial conditions, i.e. the ancestral genome and population size, were identical for all point mutation rates in our study (0.0025, 0.005, 0.0075, 0.01, 0.05, and 0.1). The avidian populations at the lowest mutation rate (0.0025) are still evolving (mean population fitness is still increasing, Supplementary Fig. S1) after 200,000 generations, explaining the higher variation in genome length for this mutation rate. Red lines are median values from 100 replicates, while the upper and lower bounds of the box are the third and first quartile, respectively. Whiskers are either 1.5 times the the quartile value or the extreme value in the data, whichever is closer to the median. Plus signs are outliers. B: The direct link between point mutation rate and genome size is further reinforced by switching the point mutation rate of population evolving at 0.0025 to 0.1 after 100,000 generations (black circles), and vice versa (green circles). The black line represents the generation where the mutation rates were switched. The long genomes shrink when mutation rate is increased and short genomes expand when mutation rate is decreased. Error bars represent $\pm$ 1 SE. Values represent the mean genome length across the population, averaged over 20 replicates.}
    \label{fig:mu_L}
\end{figure}

 We evolved avidians for 200,000 generations at six different point mutation rates. We found that genome sizes negatively correlate with the mutation rate (Fig. \ref{fig:mu_L}A; Spearman's $\rho$ = -0.72, $p < 3.6 \times 10^{-97}$). The mean population fitness also increased as the avidians' genomes grew (Supplementary Fig. S1). The point mutation rates in our experiments ranged from $2.5\times10^{-3}$ to $0.1$, and the evolved genomic mutation rates ranged from 0.13 to 24.85 (genomic mutation rate was $< 2$ for the lowest four point mutation rates). These genomic mutation rates are comparable to the ones seen in RNA viruses (0.025 in Influenza B virus, 1.1 in Hepatitis C Virus, and 4.6 in Bacteriophage Q$\beta$)~\cite{Sanjuanetal2010}. Avidians did not evolve a constant genomic mutation rate in our experiments, as Drake observed in DNA microbes and RNA viruses~\cite{DrakeHolland1999,Drake1991} and Knibbe et al reported in their digital evolution experiments~\cite{Knibbeetal2005}. A constant rate of genomic mutation is, however, not observed across the tree of life~\cite{Sniegowskietal2000}.

To test how genome size responds to changes in mutation rate, we switched the mutation rates of the avidians evolving at the lowest (0.0025) and the highest (0.1) point mutation rate after 100,000 generations. We find that the longer genomes that initially evolved at the low mutation rate began to shrink and those evolved at the high mutation rate began to expand (Fig. \ref{fig:mu_L}B), further establishing the direct influence of mutation rate on genomes size.

%talk about plos biology study that showed strong link between genome size and mutation rate
% Sloanetal2012

Since the ancestral genomes and population size were identical in all experiments, this negative correlation is independent of the effect of population size and the initial genomic content. By fixing the population size, we separated the influence of population size from that of mutation rate on genome size evolution, since it has been shown that population size influences genome size evolution as well~\cite{LynchConery2003}.

\subsection{Large genomes carry more genetic information}

Although genome size does not correlate with organismal complexity (C-value paradox), complex organisms usually do have longer genomes. In other words, while genome expansion does not necessarily increase the number of functional sites in the genome, complex organisms are likely to have a higher amount of genetic information encoded into their genomes, which requires larger genomes. For example, even though \textit{C. elegans} has a similar number of genes to \textit{H. sapiens} (19,957 genes in the nematode compared to 20,181 in humans), the nematode has 20\% less intergenic DNA and their mean intron size is 1/20th to that of humans~\cite{Taftetal2007}. On the premise that humans are more complex than \textit{C. elegans}, one can argue that the expansion of non-coding DNA is at least partly responsible for this significant increase in complexity. Indeed, \textgreater 85\% of the human genome is transcribed~\cite{Hangaueretal2013}, contributing greatly to the non-coding RNA pool of the cell that regulates expression of protein-coding genes and participate in other cellular processes~\cite{Ulitskyetal2013}. Even introns are not junk-DNA and contribute to the evolution of complexity in eukaryotes~\cite{MattickGagen2001,Rogozinetal2012}. About 20\% of the pseudogenes are transcribed in humans~\cite{Zhengetal2007}, and are differentially expressed in cancers and viral infections~\cite{Polisenoetal2010,Guptaetal2015}. Thus, genome expansion, even if primarily of the non-coding DNA, likely increases the number of functional sites in the genome. Even if some of this inserted DNA is non-functional at the outset, evolution can repurpose it to achieve higher organismal complexity and genetic information~\cite{Lynch2007,Knibbeetal2007}.

\begin{figure}[!h]
%    \centering
    \includegraphics[width=1.0\textwidth]{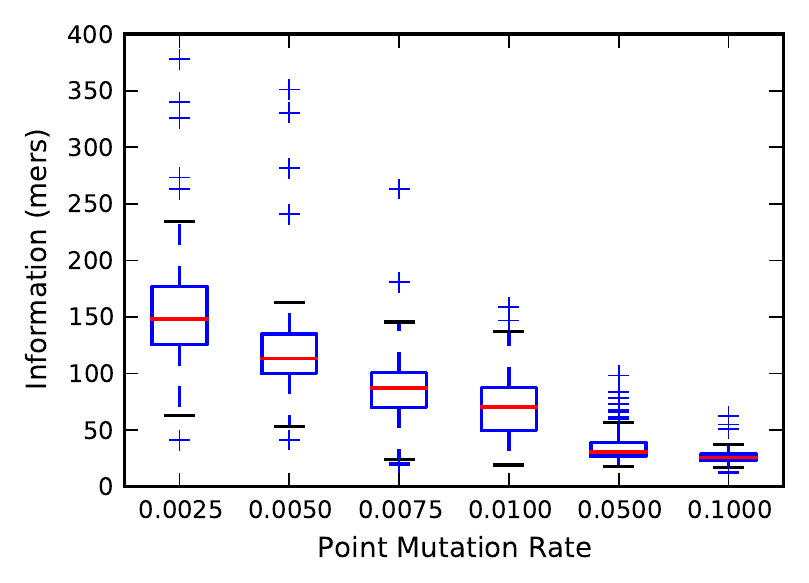}
    \caption{Genomes evolved at low mutation rates had higher genetic information (number of essential sites in the genome, see Methods) than genomes evolved at high mutation rates. The information measure is reported for the fittest genotype in each of the 100 replicate populations. Red lines are median values from 100 replicates, while the upper and lower bounds of the box are the third and first quartile, respectively. Whiskers are either 1.5 times the the quartile value or the extreme value in the data, whichever is closer to the median. Plus signs are outliers.}
    \label{fig:info_tasks}
\end{figure}

In our experiments, avidians that evolved long genomes at low mutation rates had higher genetic information (number of essential sites in the genome) than those that evolved at high mutation rates and had shorter genomes (Fig. \ref{fig:info_tasks}; Spearman's $\rho$ = -0.86, $p < 6.4 \times 10^{-180}$). The longer genomes also evolved more traits (see Methods for an explanation of traits, and Supplementary Fig. S2), which are the computational equivalent of biological pathways that lead to observable phenotypes. The mean population fitness was also inversely related to mutation rate, although the mean fitness of populations evolving at point mutation rate of 0.0025 was still increasing after 200,000 generations (Supplementary Fig. S1). This suggests that larger genome size is a necessary, if not sufficient, requirement for evolving phenotypic novelty. The avidians on average evolved fewer traits when the point mutation rate was switched half-way from 0.0025 to 0.1, and evolved more traits when mutation rate was switched from 0.1 to 0.0025, emphasizing the relationship between genome size, mutation rate, and phenotypic complexity (Fig. \ref{fig:mu_L}B and Supplementary Fig. S3).

\subsection{Beneficial insertions drive genome expansion at low mutation rates}

\begin{figure}[!htb]
%    \centering
    \includegraphics[width=\textwidth]{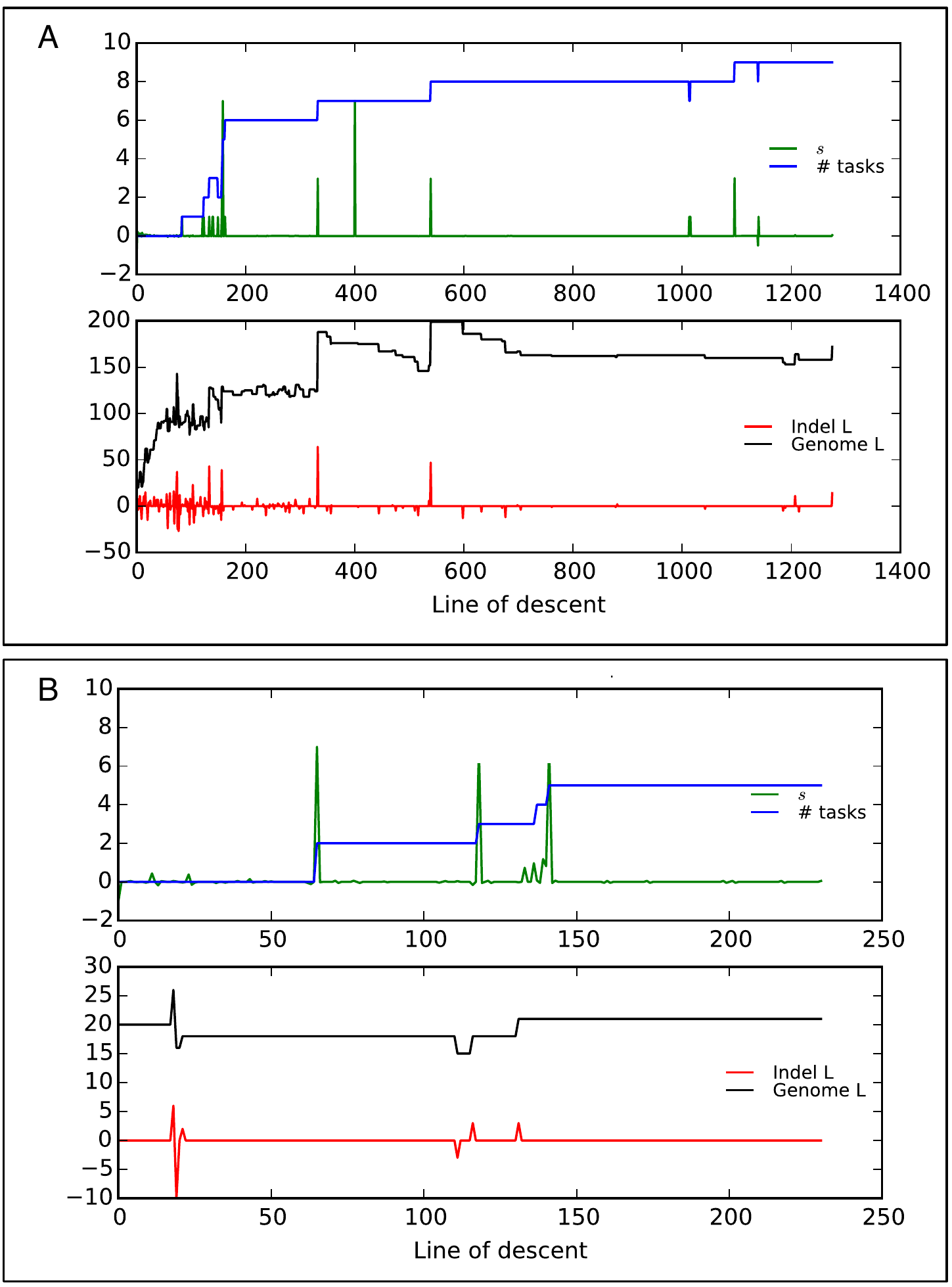}
    \caption{The line of descent (LOD) of the most fit genome is shown for a single replicate population evolving at the lowest (0.0025, A) and the highest (0.1, B) point mutation rate in our study. The fitness effects of genome edit events (insertions, deletions, base substitutions) are shown in green, the number of evolved traits is shown in blue, the size of indels is shown in red, and the genome length is shown in black. At low mutation rate (top panel, A), new traits (in blue) often evolved following beneficial genomic events (green spikes), and are sometimes concurrent with insertion events (red spikes). These beneficial insertions appear to increase the genome size (black line) over time. At the high mutation rate (bottom panel, B), insertion events are not as frequent as at low mutation rates (also see Supplementary Fig. S4), with genome size staying relatively constant. The line of descent (LOD) maps for other mutation rates can be found in Supplementary Fig. S5.}
    \label{fig:LOD}
\end{figure}

\begin{figure}[!htb]
%    \centering
    \includegraphics[width=\textwidth]{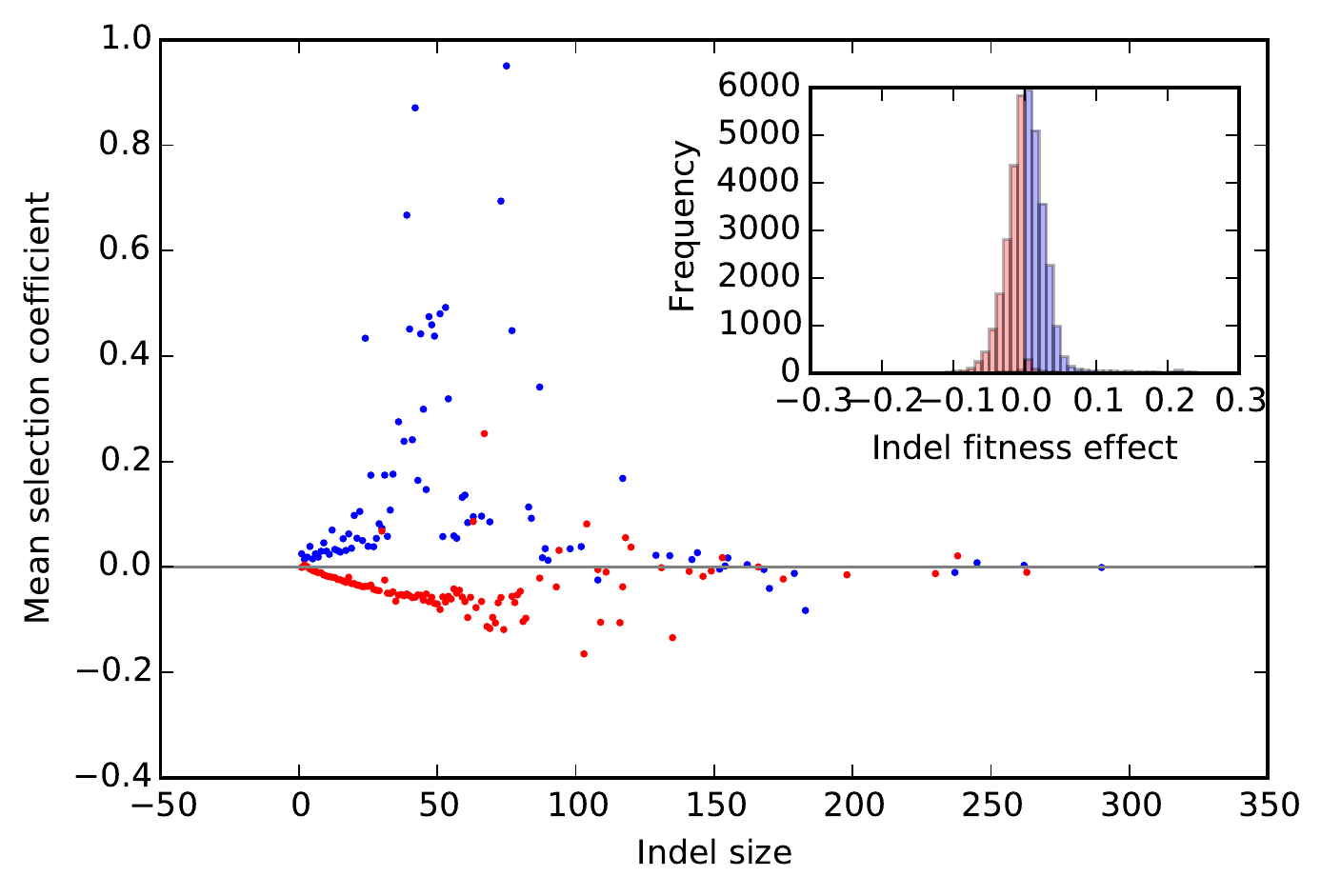}
    \caption{The average fitness effect of insertions (blue) and deletions (red) as a function of indel size is shown for 100 replicate populations evolving at the point mutation rate of 0.0025. Indels above the gray line ($s$ = 0) are beneficial and those below the gray line are deleterious. Small insertions (blue dots) are usually beneficial, while small deletions (red dots) are usually deleterious. The inset plot shows the histograms of fitness effects of insertions (blue bars, total 19,262 insertions) and deletions (red bars, total 16,998 deletions) along the line of descent in 100 replicate populations. Insertions (blue bars) are usually beneficial (\textit{i.e.}, fitness effect \textgreater 0), and deletions (red bars) are usually deleterious (fitness effect \textless 0). The two distributions are significantly different (Kolmogorov-Smirnov two-sided test, $p < 1\times10^{-100}$).}
    \label{fig:indel_tasks}
\end{figure}

To understand how genomes gain meaningful increases in size, we followed the genome edits (indels and mutations), the corresponding effect on fitness ($s$), the number of traits evolved, and genome size along the line of descent in avidians evolving at different mutation rates (Fig. \ref{fig:LOD}). At the lowest point mutation rate in our experiments (Fig. \ref{fig:LOD}A), the beneficial changes in the genome (green spikes) often align with evolution of new traits (blue line), as well as with insertions in the genome (red spikes). Insertions are largely beneficial compared to deletions at low mutation rate (Fig. \ref{fig:indel_tasks}). Phenotypic innovation (evolving a new trait) was preceded by insertion events 87\% of the time (within the previous 20 ancestors along the line of descent), while deletions preceded innovation 60\% of the time (null hypothesis: presence or absence of insertions is irrelevant to trait evolution, rejected with $ p < 1.0\times10^{-100}$, ${\chi}^2$ test statistic = $2.23\times10^{5}$). Thus, avidian genomes are likely to evolve new traits after an insertion event, suggesting that phenotypic innovation happens in a two-step process: genome expansion followed by evolution of a new trait by substitutions. Insertions are not deleterious per se (inset plots in Fig. \ref{fig:indel_tasks}) and thus persist in the line of descent. In fact, these inserted sequences may serve as substrates for evolving new phenotypic traits later on, contributing to increase in fitness and phenotypic complexity. In contrast, indels are infrequent at high mutation rates on the line of descent (Fig. \ref{fig:LOD}B, also see Supplementary Fig. S4). As a result, the genomes do not grow and evolve fewer traits compared to the genomes evolved at low mutation rates.

%\newpage
%\newpage

This prominent role of beneficial insertions in genome evolution of asexual organisms is in contrast to how genome sizes are shaped by DNA loss in eukaryotes. The reported biases in indel spectra (rarity of long insertions and abundance of short deletions) are seen primarily in eukaryotic genomes~\cite{Petrov2002}. Yet, a thermodynamic argument suggests that large indels are likely to increase genome size, since insertion events require only one breakpoint in the genome rendering large insertions less disruptive than large deletions~\cite{Petrov2002,Gregory2004}. By the same argument, DNA loss is more likely to happen by small deletions to minimize the fitness cost to the organism. Thus, while eukaryotic genomes may evolve by rapid expansion due to whole genome duplication events and TE proliferation, asexual populations such as RNA viruses may have grown their genomes gradually via beneficial insertions. However, gradual increases in avidian genomes at low mutation rates is still followed by small deletions that fine-tune the genome size (Fig. \ref{fig:LOD}).

%\newpage

%\newpage
%talk about how bacteria (a prokaryote) fits into this picture, or if our studies are more in line with rna viruses than viruses and prokaryotes. bacteria have some transposable activity as well as horizontal gene transfer activity

\subsection{High mutation rates force genomes to be small and informationally dense}

\begin{figure}[!h]
%    \centering
    \includegraphics[width=1.0\textwidth]{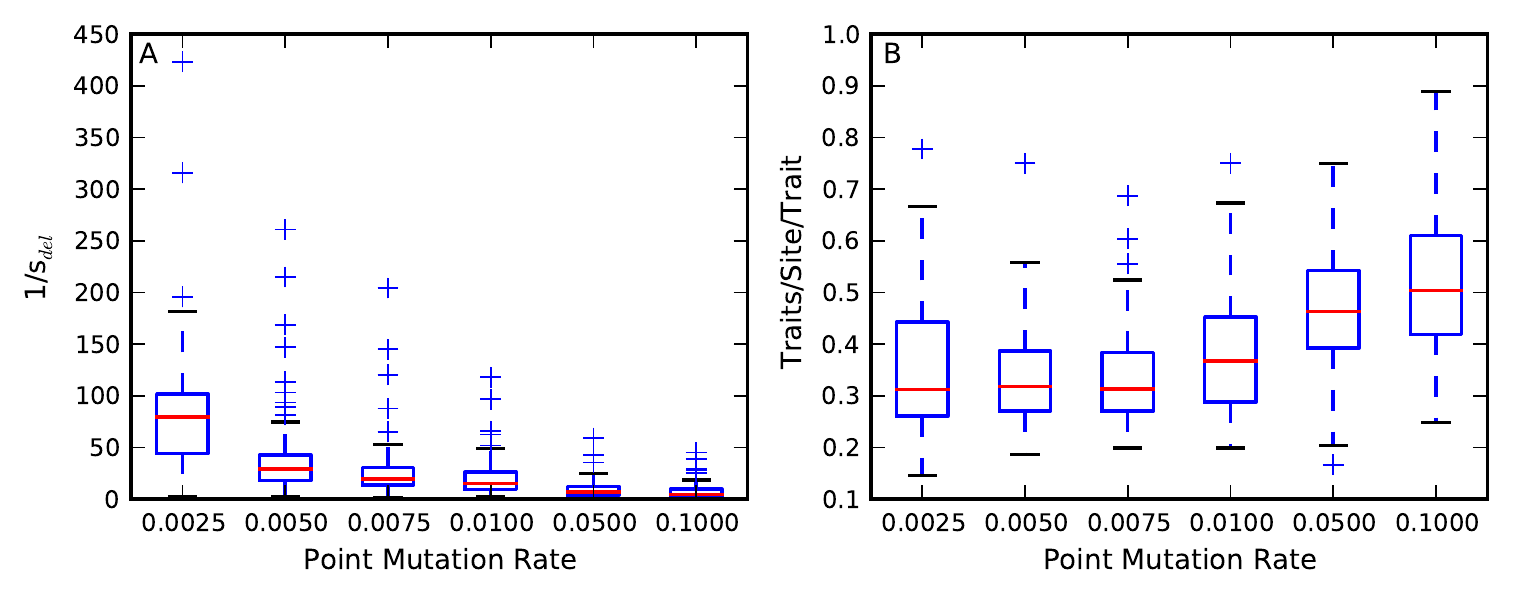}
    \caption{Deleterious mutations at high mutations rates are more costly due to informationally dense genomes. The inverse of the harmonic mean of deleterious selection coefficients for the fittest genotype from each replicate shows that deleterious mutations are costlier at high mutation rates (A). This can be explained by the high coding density in these genomes (B). Traits/site/trait represents how many traits are encoded per site, normalized by the total number of evolved traits, and thus is a measure of coding density of the genome. Red lines are median values from 100 replicates, while the upper and lower bounds of the box are the third and first quartile, respectively. Whiskers are either 1.5 times the the quartile value or the extreme value in the data, whichever is closer to the median. Plus signs are outliers.}
    \label{fig:robustness_density}
\end{figure}

If beneficial insertions drive genome expansion at low mutation rates, what keeps genomes small at high mutation rates? We find that the fitness cost of deleterious mutations is high at high mutation rates (Fig. \ref{fig:robustness_density}A; Spearman's $\rho$ = -0.71, $p < 1.1 \times 10^{-90}$). Since genotypes evolving at high mutation rates are compact, genetic information is forced to be distributed over a small number of sites (Fig. \ref{fig:robustness_density}B), as in overlapping genes commonly seen in viral genomes. A deleterious mutation at a single such site can unfavorably affect multiple traits, increasing the overall fitness cost of deleterious mutations. Digital evolution experiments also find that gene knockouts are more deleterious when pleiotropy is high, as is common in compact genomes~\cite{Knibbeetal2007b}. Thus, not only is the mutational load high at high mutation rates, the deleterious mutations are costlier than they are at low mutation rates (mutational load is $1 - e^{-\mu s}$, as derived from~\cite{KimuraMaruyama1966}). This compounding factor only strengthens the selection pressure to decrease mutational load by reducing genome size, especially since population size is fixed in our experiments.

It should be noted that mutation rate can itself evolve to facilitate adaptation (reviewed in~\cite{Sniegowskietal2000} and~\cite{MetzgarWills2000}). For example, the mutator strain of \textit{E. coli} with a higher mutation rate than the wild-type bacteria showed the ability to adapt faster~\cite{Giraudetal2001}. Even though the  majority of mutations are deleterious, the ability to quickly find the adaptive beneficial mutations was enough to increase the population of the mutator strain relative to the wild-type~\cite{Giraudetal2001}. However, this evolutionary advantage is short-lived and disappears once the beneficial mutations are found and there is no more fitness peak to climb~\cite{Giraudetal2001,Arjanetal1999}. The mutator strain also does not propagate faster than the wild-type when a higher mutation supply is achieved by increasing the population size~\cite{Giraudetal2001,Arjanetal1999}. 
%These observations suggest that mutation rate, genome size, and population size are parameters that evolution optimizes to achieve the mutation supply rate necessary for adaption. 
Thus, environmental stresses such as starvation triggers a response in bacteria wherein mutation rate is elevated to quickly find beneficial mutations to adapt to the temporarily adverse conditions~\cite{Rosenbergetal1998}. %,Foster2000}.

Since high mutation rate increases the mutational load in an evolving population, it makes sense that when the environmental stress is no longer present, the mutation rate would revert to the lower level. After all, the fitness cost of accumulating deleterious mutations would be too high if the rapid rate of adaptation afforded by high mutation rate is not needed. Mutator strains in well-adapted bacterial populations evolve decreased mutation rate as the opportunity for adaptation diminishes~\cite{Wielgossetal2013}, an observation supported by digital evolution experiments~\cite{Cluneetal2008}. Perhaps a continual need for adaptation is responsible for consistently high mutation rates in viruses, parasites, and sometimes in pathogenic bacteria where rapid adaptation to host immune responses is critical for surviving such an evolutionary arms race~\cite{GreenspoonMGonigle2013,Sniegowskietal2000,Hobothetal2009,ElenaSanjun2005}. The selection pressure to adapt quickly to a changing environment appears to trump the selection pressure to decrease mutational load by minimizing the mutation rate. However, mutational load can restrict virus adaptability due to an abundance of deleterious mutations~\cite{Pybusetal2007}. Thus, the compromise between evolutionary forces for reducing the mutation load and maintaining high adaptability might shape the genome size and information density in RNA viruses.

% find the lineage step by which the first task is learned (rate of adaptation)

% <I think for a robustness argument, we will have to look at fraction of sites that are neutral as well as lethal. Do we need to make a robustness argument? I'm not so sure. Also, these data are from the most fit genomes in each population?>

%\newpage
%\newpage

%<talk about fixed population size- how population size is also linked to inverse L>
%<talk about plos biology paper, experimentally confirmed that mutation rate is linked to genome size, independent of population size; see what mechanism they propose: TE elments or indels? Their system is eukaryotic and genomes are mitochondrial. We generalize thir findings. They also find increase in complexity in genomic architecture, we find increase in information - draw parallels and then talks about indel spectra effect that we find>

%<talk about coding density paper: how we also find high coding density in high mu rate>
%<talk about mutational robustness in RNA viruses: asexual populations (with exceptions) at high mutation rates>

\section{Conclusions}

While empirical studies reveal significant aspects of genome size evolution, digital evolution systems provide an opportunity to observe evolution-in-action and to manipulate evolutionary parameters in ways that allows exploring the relative importance of the many evolutionary forces that simultaneously act on genomes. Comparative genomics analyses have unearthed important relationships between population size, mutation rate, gene content, genome size, and their combined influence on evolution of complexity. However, digital evolution experiments complement these retrospective observations by investigating evolutionary processes that are difficult to test experimentally. In our experiments at a range of mutation rates, we find concurrence with the empirical finding that the point mutation rate is negatively correlated with genome size. By tracking the genomes along the line of descent, we find insertions to be significantly beneficial compared to deletions, suggesting that before the advent of complex mechanisms of genome edits such as TE activity, beneficial insertions drove genome expansion. That these insertions are followed by phenotypic innovations further explains why insertions are evolutionarily favored in asexual populations. At the same time, the point mutation rate influences genome size via the mutational load. Thus, unless high mutation rate provides a critical evolutionary advantage such as rapid adaption to a temporary environmental stress, the selection pressure to reduce mutational load forces the genomes to shrink at high mutation rates. This genome shrinkage results in genomes packed with genetic information, and this compactness likely increases the fitness cost of deleterious mutations, further compounding the severity of mutational load. Still, a high point mutation rate is frequently seen in natural populations, especially in viruses, suggesting that the selection pressure to maintain high evolvability (for example, against a highly adaptive host immune system) can take precedence over selection pressure to reduce mutational load in the fight to survival.

The evolution of genome size is a complex phenomenon, especially in eukaryotes due to TE activity and expansion of non-coding DNA. Our analyses of asexual populations evolving at fixed point mutation and indel rates reveal the fundamental roles that indel spectra and mutational load play in determining genome size and phenotypic diversity. Investigations into eukaryotic genome size evolution by including recombination and TE activity in digital evolution platforms will allow comparisons with asexual genome size evolution, and can shed light on evolution of complex genome editing mechanisms.

\section{Materials and Methods}
\subsection{Avida digital evolution platform} 

Avida is a digital evolution platform which provides an environment within which digital organisms, using sets of instructions analogous to codons, experience selective pressures to develop genes that encode logical operations~\cite{Adamietal1994,Adami1998,OfriaWilke2004}. Performing these operations provides these “avidians” with “single instruction processing units” or SIPs, their energy currency equivalent of ATP. By performing increasingly complex boolean logic calculations, the avidians are able to accrue larger amounts of energy to outcompete their neighbors, much as living organisms participate in the evolutionary arms race in their own ecological niches. They replicate by error-prone mechanisms, thus mimicking Darwinian evolution. Since we investigated qualities that are innate to genomes as stores of information, and are not mechanism dependent (other than a requirement for a lack of total fidelity in replication), Avida is an ideal model system to study evolutionary forces that drive genome evolution in asexual populations. 

\subsection{Experimental Design}
To test the role of the mutation rate in driving genome size evolution, we evolved 100 replicate populations at various point mutation rates ($\mu$ = $\{$0.0025, 0.005, 0.0075, 0.01, 0.05, 0.1$\}$) for 200,000 generations. Insertions and deletions occurred with equal frequency at a constant rate of 0.05 indels per generation. Indel size was uniformly distributed, with genome size changing at most by 10\% in any given generation. All populations were initialized with an identical ancestral genome of size 20. Population size was fixed at the default 3600 individuals. There was no structure in the evolving populations (i.e. a well-mixed environment). An additional 40 populations were evolved for 200,000 generations where the mutation rates were switched after 100,000 generations as follows: 20 populations that initially evolved at a point mutation rate of 0.0025 were switched to a point mutation rate of 0.1 after 100,000 generations, and the remaining 20 populations were switched from point mutation rate of 0.1 to 0.0025 after 100,000 generations.

\subsection{Line of Descent}
To track the effect of genome edits on genome size and phenotypic evolution, we analyzed the Line of Descent (LOD) of the fittest individual from each replicate population at the end of the evolution experiments. A LOD is a lineage of every ancestor of the evolved genotype that had the highest fitness at the end of 200,000 generations. It tracks every genome edit (and its corresponding effect on fitness) that was fixed in the lineage. This genotypic ``fossil record" allows identifying those mutations that lead to evolutionary innovations and determine the respective role of insertions and deletions in genome size evolution.  

\subsection{Data Analysis}
We calculated statistics at both the population level and for individual genotypes. The mean genome length and the mean fitness was calculated by averaging the relevant values across all genotypes in each population which was then averaged over 100 replicate populations. For the rest of our reported data, we calculated statistics from the fittest genotype in the final evolved population. A genotype's information content was estimated as $I = L -\sum_{i}^{L}\log_{26}\nu(i)$, where L is the genome size, 26 is the alphabet size for avidian genomes, and $\nu(i)$ is the number of mutations that are neutral or beneficial (see~\cite{Adamietal2000} for further explanation of this estimation). Thus, information content is a measure of the number of essential sites in a genome. The number of phenotypic traits a genotype possesses is calculated as the number of different boolean logic calculations it can perform. 

The traits per site per trait measure is determined by performing knockout mutations at every site in the genome and then counting the number of traits that are lost due to each knockout mutation (lethal knockouts are not considered). This gives the number of traits that utilize each genomic site, and average of this quantity over the length of the genome gives the overall number of traits encoded per site. The normalized trait/site/trait is then calculated by dividing the traits/site by the total number of traits evolved by the genome. 

%Four populations yielded fittest genotypes that were non-viable at the end of 200,000 generations (out of 600 populations: 100 replicates for each of the 6 point mutation rates), and were excluded from the analyses.

\section{Acknowledgments}
This work was supported in part by the National Science Foundation's BEACON Center for the Study of Evolution in Action, under contract No. DBI-0939454. We wish to acknowledge the Michigan State University High Performance Computing Center and the Institute for Cyber Enabled Research for computational support. Michael Miyagi thanks the Freshman Research Initiative program at the University of Texas at Austin for undergraduate research opportunity.

\bibliographystyle{unsrt}

\begin{thebibliography}{99}

\bibitem{Taftetal2007}
Ryan~J Taft, Michael Pheasant, and John~S Mattick.
\newblock The relationship between non-protein-coding dna and eukaryotic
  complexity.
\newblock {\em Bioessays}, 29(3):288--99, Mar 2007.

\bibitem{LynchConery2003}
Michael Lynch and John~S Conery.
\newblock The origins of genome complexity.
\newblock {\em Science}, 302(5649):1401--4, Nov 2003.

\bibitem{Sniegowskietal2000}
P~D Sniegowski, P~J Gerrish, T~Johnson, and A~Shaver.
\newblock The evolution of mutation rates: separating causes from consequences.
\newblock {\em Bioessays}, 22(12):1057--66, Dec 2000.

\bibitem{Lynch2010}
Michael Lynch.
\newblock Evolution of the mutation rate.
\newblock {\em Trends Genet}, 26(8):345--52, Aug 2010.

\bibitem{Holmes2003}
Edward~C Holmes.
\newblock Error thresholds and the constraints to rna virus evolution.
\newblock {\em Trends Microbiol}, 11(12):543--6, Dec 2003.

\bibitem{Zwartetal2014}
Mark~P Zwart, Anouk Willemsen, Jos{\'e}-Antonio Dar{\`o}s, and Santiago~F
  Elena.
\newblock Experimental evolution of pseudogenization and gene loss in a plant
  rna virus.
\newblock {\em Mol Biol Evol}, 31(1):121--34, Jan 2014.

\bibitem{Tromasetal2014}
Nicolas Tromas, Mark~P Zwart, Javier Forment, and Santiago~F Elena.
\newblock Shrinkage of genome size in a plant rna virus upon transfer of an
  essential viral gene into the host genome.
\newblock {\em Genome Biol Evol}, 6(3):538--50, Mar 2014.

\bibitem{Kuoetal2009}
Chih-Horng Kuo, Nancy~A Moran, and Howard Ochman.
\newblock The consequences of genetic drift for bacterial genome complexity.
\newblock {\em Genome Res}, 19(8):1450--4, Aug 2009.

\bibitem{McCutcheonMoran2012}
John~P McCutcheon and Nancy~A Moran.
\newblock Extreme genome reduction in symbiotic bacteria.
\newblock {\em Nat Rev Microbiol}, 10(1):13--26, Jan 2012.

\bibitem{Lynch2006}
Michael Lynch.
\newblock Streamlining and simplification of microbial genome architecture.
\newblock {\em Annu Rev Microbiol}, 60:327--49, 2006.

\bibitem{Vinogradov2004}
Alexander~E Vinogradov.
\newblock Evolution of genome size: multilevel selection, mutation bias or
  dynamical chaos?
\newblock {\em Curr Opin Genet Dev}, 14(6):620--6, Dec 2004.

\bibitem{Petrovetal2000}
D~A Petrov, T~A Sangster, J~S Johnston, D~L Hartl, and K~L Shaw.
\newblock Evidence for dna loss as a determinant of genome size.
\newblock {\em Science}, 287(5455):1060--2, Feb 2000.

\bibitem{Gregory2004}
T~Ryan Gregory.
\newblock Insertion-deletion biases and the evolution of genome size.
\newblock {\em Gene}, 324:15--34, Jan 2004.

\bibitem{KuoOchman2009}
Chih-Horng Kuo and Howard Ochman.
\newblock Deletional bias across the three domains of life.
\newblock {\em Genome Biol Evol}, 1:145--52, 2009.

\bibitem{OfriaWilke2004}
Charles Ofria and Claus~O Wilke.
\newblock Avida: a software platform for research in computational evolutionary
  biology.
\newblock {\em Artif Life}, 10(2):191--229, 2004.

\bibitem{Adamietal2000}
C~Adami, C~Ofria, and T~C Collier.
\newblock Evolution of biological complexity.
\newblock {\em Proc Natl Acad Sci U S A}, 97(9):4463--8, Apr 2000.

\bibitem{Batutetal2013}
B{\'e}r{\'e}nice Batut, David~P Parsons, Stephan Fischer, Guillaume Beslon, and
  Carole Knibbe.
\newblock In silico experimental evolution: a tool to test evolutionary
  scenarios.
\newblock {\em BMC Bioinformatics}, 14 Suppl 15:S11, 2013.

\bibitem{Lenskietal2003}
Richard~E Lenski, Charles Ofria, Robert~T Pennock, and Christoph Adami.
\newblock The evolutionary origin of complex features.
\newblock {\em Nature}, 423(6936):139--44, May 2003.

\bibitem{Wilkeetal2001}
C~O Wilke, J~L Wang, C~Ofria, R~E Lenski, and C~Adami.
\newblock Evolution of digital organisms at high mutation rates leads to
  survival of the flattest.
\newblock {\em Nature}, 412(6844):331--3, Jul 2001.

\bibitem{Zamanetal2014}
Luis Zaman, Justin~R Meyer, Suhas Devangam, David~M Bryson, Richard~E Lenski,
  and Charles Ofria.
\newblock Coevolution drives the emergence of complex traits and promotes
  evolvability.
\newblock {\em PLoS Biol}, 12(12):e1002023, Dec 2014.

\bibitem{Elenaetal2008}
Santiago~F Elena and Rafael Sanju{\'a}n.
\newblock The effect of genetic robustness on evolvability in digital
  organisms.
\newblock {\em BMC Evol Biol}, 8:284, 2008.

\bibitem{ODonnelletal2014}
Daniel~R O'Donnell, Abhijna Parigi, Jordan~A Fish, Ian Dworkin, and Aaron~P
  Wagner.
\newblock The roles of standing genetic variation and evolutionary history in
  determining the evolvability of anti-predator strategies.
\newblock {\em PLoS One}, 9(6):e100163, 2014.

\bibitem{Sanjuanetal2010}
Rafael Sanju{\'a}n, Miguel~R Nebot, Nicola Chirico, Louis~M Mansky, and Robert
  Belshaw.
\newblock Viral mutation rates.
\newblock {\em J Virol}, 84(19):9733--48, Oct 2010.

\bibitem{DrakeHolland1999}
J~W Drake and J~J Holland.
\newblock Mutation rates among rna viruses.
\newblock {\em Proc Natl Acad Sci U S A}, 96(24):13910--3, Nov 1999.

\bibitem{Drake1991}
J~W Drake.
\newblock A constant rate of spontaneous mutation in dna-based microbes.
\newblock {\em Proc Natl Acad Sci U S A}, 88(16):7160--4, Aug 1991.

\bibitem{Knibbeetal2005}
Carole Knibbe, Guillaume Beslon, Virginie Lefort, F~Chaudier, and J-M Fayard.
\newblock Self-adaptation of genome size in artificial organisms.
\newblock In {\em Advances in Artificial Life}, pages 423--432. Springer, 2005.

\bibitem{Hangaueretal2013}
Matthew~J Hangauer, Ian~W Vaughn, and Michael~T McManus.
\newblock Pervasive transcription of the human genome produces thousands of
  previously unidentified long intergenic noncoding rnas.
\newblock {\em PLoS Genet}, 9(6):e1003569, Jun 2013.

\bibitem{Ulitskyetal2013}
Igor Ulitsky and David~P Bartel.
\newblock lincrnas: genomics, evolution, and mechanisms.
\newblock {\em Cell}, 154(1):26--46, Jul 2013.

\bibitem{MattickGagen2001}
J~S Mattick and M~J Gagen.
\newblock The evolution of controlled multitasked gene networks: the role of
  introns and other noncoding rnas in the development of complex organisms.
\newblock {\em Mol Biol Evol}, 18(9):1611--30, Sep 2001.

\bibitem{Rogozinetal2012}
Igor~B Rogozin, Liran Carmel, Miklos Csuros, and Eugene~V Koonin.
\newblock Origin and evolution of spliceosomal introns.
\newblock {\em Biol Direct}, 7:11, 2012.

\bibitem{Zhengetal2007}
Deyou Zheng, Adam Frankish, Robert Baertsch, Philipp Kapranov, Alexandre
  Reymond, Siew~Woh Choo, Yontao Lu, France Denoeud, Stylianos~E Antonarakis,
  Michael Snyder, Yijun Ruan, Chia-Lin Wei, Thomas~R Gingeras, Roderic
  Guig{\'o}, Jennifer Harrow, and Mark~B Gerstein.
\newblock Pseudogenes in the encode regions: consensus annotation, analysis of
  transcription, and evolution.
\newblock {\em Genome Res}, 17(6):839--51, Jun 2007.

\bibitem{Polisenoetal2010}
Laura Poliseno, Leonardo Salmena, Jiangwen Zhang, Brett Carver, William~J
  Haveman, and Pier~Paolo Pandolfi.
\newblock A coding-independent function of gene and pseudogene mrnas regulates
  tumour biology.
\newblock {\em Nature}, 465(7301):1033--8, Jun 2010.

\bibitem{Guptaetal2015}
Aditi Gupta, C.~Titus Brown, Yong-Hui Zheng, and Christoph Adami.
\newblock Differentially-expressed pseudogenes in hiv-1 infection.
\newblock {\em Viruses}, 7(10):5191--5205, 2015.

\bibitem{Lynch2007}
Michael Lynch.
\newblock The frailty of adaptive hypotheses for the origins of organismal
  complexity.
\newblock {\em Proc Natl Acad Sci U S A}, 104 Suppl 1:8597--604, May 2007.

\bibitem{Knibbeetal2007}
Carole Knibbe, Antoine Coulon, Olivier Mazet, Jean-Michel Fayard, and Guillaume
  Beslon.
\newblock A long-term evolutionary pressure on the amount of noncoding dna.
\newblock {\em Mol Biol Evol}, 24(10):2344--53, Oct 2007.

\bibitem{Petrov2002}
Dmitri~A Petrov.
\newblock Mutational equilibrium model of genome size evolution.
\newblock {\em Theor Popul Biol}, 61(4):531--44, Jun 2002.

\bibitem{Knibbeetal2007b}
Carole Knibbe, Olivier Mazet, Fabien Chaudier, Jean-Michel Fayard, and
  Guillaume Beslon.
\newblock Evolutionary coupling between the deleteriousness of gene mutations
  and the amount of non-coding sequences.
\newblock {\em J Theor Biol}, 244(4):621--30, Feb 2007.

\bibitem{KimuraMaruyama1966}
M~Kimura and T~Maruyama.
\newblock The mutational load with epistatic gene interactions in fitness.
\newblock {\em Genetics}, 54(6):1337--51, Dec 1966.

\bibitem{MetzgarWills2000}
D~Metzgar and C~Wills.
\newblock Evidence for the adaptive evolution of mutation rates.
\newblock {\em Cell}, 101(6):581--4, Jun 2000.

\bibitem{Giraudetal2001}
A~Giraud, I~Matic, O~Tenaillon, A~Clara, M~Radman, M~Fons, and F~Taddei.
\newblock Costs and benefits of high mutation rates: adaptive evolution of
  bacteria in the mouse gut.
\newblock {\em Science}, 291(5513):2606--8, Mar 2001.

\bibitem{Arjanetal1999}
J~A Arjan, M~Visser, C~W Zeyl, P~J Gerrish, J~L Blanchard, and R~E Lenski.
\newblock Diminishing returns from mutation supply rate in asexual populations.
\newblock {\em Science}, 283(5400):404--6, Jan 1999.

\bibitem{Rosenbergetal1998}
S~M Rosenberg, C~Thulin, and R~S Harris.
\newblock Transient and heritable mutators in adaptive evolution in the lab and
  in nature.
\newblock {\em Genetics}, 148(4):1559--66, Apr 1998.

\bibitem{Wielgossetal2013}
S{\'e}bastien Wielgoss, Jeffrey~E Barrick, Olivier Tenaillon, Michael~J Wiser,
  W~James Dittmar, St{\'e}phane Cruveiller, B{\'e}atrice Chane-Woon-Ming,
  Claudine M{\'e}digue, Richard~E Lenski, and Dominique Schneider.
\newblock Mutation rate dynamics in a bacterial population reflect tension
  between adaptation and genetic load.
\newblock {\em Proc Natl Acad Sci U S A}, 110(1):222--7, Jan 2013.

\bibitem{Cluneetal2008}
Jeff Clune, Dusan Misevic, Charles Ofria, Richard~E Lenski, Santiago~F Elena,
  and Rafael Sanju{\'a}n.
\newblock Natural selection fails to optimize mutation rates for long-term
  adaptation on rugged fitness landscapes.
\newblock {\em PLoS Comput Biol}, 4(9):e1000187, 2008.

\bibitem{GreenspoonMGonigle2013}
Philip~B Greenspoon and Leithen~K M'Gonigle.
\newblock The evolution of mutation rate in an antagonistic coevolutionary
  model with maternal transmission of parasites.
\newblock {\em Proc Biol Sci}, 280(1761):20130647, Jun 2013.

\bibitem{Hobothetal2009}
Christina Hoboth, Reinhard Hoffmann, Anja Eichner, Christine Henke, Sabine
  Schmoldt, Axel Imhof, J{\"u}rgen Heesemann, and Michael Hogardt.
\newblock Dynamics of adaptive microevolution of hypermutable pseudomonas
  aeruginosa during chronic pulmonary infection in patients with cystic
  fibrosis.
\newblock {\em J Infect Dis}, 200(1):118--30, Jul 2009.

\bibitem{ElenaSanjun2005}
Santiago~F Elena and Rafael Sanju{\'a}n.
\newblock Adaptive value of high mutation rates of rna viruses: separating
  causes from consequences.
\newblock {\em J Virol}, 79(18):11555--8, Sep 2005.

\bibitem{Pybusetal2007}
Oliver~G Pybus, Andrew Rambaut, Robert Belshaw, Robert~P Freckleton, Alexei~J
  Drummond, and Edward~C Holmes.
\newblock Phylogenetic evidence for deleterious mutation load in rna viruses
  and its contribution to viral evolution.
\newblock {\em Mol Biol Evol}, 24(3):845--52, Mar 2007.

\bibitem{Adamietal1994}
Chris Adami and C.~Titus Brown.
\newblock Evolutionary learning in the 2d artificial life system `avida'.
\newblock {\em Artificial life IV.}, 1194:377--381, 1994.

\bibitem{Adami1998}
Christoph. Adami.
\newblock {\em Introduction to artificial life}, volume~1.
\newblock Springer Science \& Business Media, 1998.

\end{thebibliography}

\newpage

\section*{Supplementary Materials}
\setcounter{equation}{0}
\setcounter{figure}{0}
\setcounter{table}{0}
\setcounter{page}{1}
\makeatletter
\renewcommand{\theequation}{S\arabic{equation}}
\renewcommand{\thefigure}{S\arabic{figure}}

\subsection*{Expanded Methods: Avida}
Avida is a digital experimental evolution platform where populations of simple computer programs (avidians) compete for the resources needed to self-replicate via error-prone mechanisms. The avidian genome consists of computer instructions which are executed during its life cycle to perform boolean logic calculations as well as to replicate its genome. Since evolution in Avida comprises genetic variation affecting ability to evolve phenotypic traits and to replicate, differential fitness dependent on this heritable variation and competition for computational resources causes avidians to undergo natural selection comparable to biological populations.
%Since evolution in Avida comprises genetic variation affecting ability to perform tasks and to replicate, differential fitness dependent on this heritable variation, and competition for computational resources, these computer programs, called avidians, undergo evolution by natural selection comparable to biological populations. 

The Avida world consists of a 60x60 toroidal grid with at most one avidian per cell, resulting in a fixed population size of 3600. Each child avidian is placed in any one of the 3600 cells after successful replication (although new offspring are preferentially placed in empty cells if available), making the population well-mixed. When the population is at its carrying capacity, the avidian occupying the cell chosen for a new offspring will be removed from the population. This random selection of individuals for removal adds an element of genetic drift to avidian populations.

% I think this detailed discussion of time in avida is unnecessary. Also, is reproduction speed is the only target? and not fitness gain by learning tasks?

Absolute time in Avida is divided into updates. During each update, the population executes 30N instructions, where N is the population size. The ability to execute these instructions (comparable to energy units in cells--ATP), called Single Instruction Processing Units (SIPs), are distributed across the population. How these SIPs get distributed among the individuals in the population is dependent on a characteristic possessed by each individual called merit. 
%A genotype's relative merit determines how many SIPs it will receive every update. 
In a monoclonal population, every individual will possess on average 30 SIPs per update. However, if one individual has a greater merit than others in the population, it is expected to receive more SIPs per update than the other individuals. This allows it to execute and copy its genome faster than other individuals. Therefore, as reproduction speed is the primary target of selection in this type of simple environment, increased merit results in increased fitness, and organisms with an increased merit will be under positive selection. In our experiments, we record data every generation, starting from the ancestral population, which marks generation 0. All progeny of the ancestral population constitute generation 1, and so forth.

Avidians increase their merit through the evolution of phenotypic traits. These traits are the ability to perform boolean logic computations. In the default Avida environment, the Logic-9 environment~\cite{Lenskietal2003}, populations can evolve up to 9 of these traits. Performing these traits result in a multiplicative increase in an individual's merit (ranging from a multiple of 2 for simple traits to 32 for the most complex trait). The evolution of these traits require many point mutations and a genome size large enough to contain the instructions necessary to perform these computations. Because these traits increase merit, and thus replication speed, the evolution of these traits are also under strong selection. Each individual can perform each trait once during their lifespan, and there is no limit to the number of times a trait can be performed in a population. Because an individual's performance of a trait does not limit the others in the population, there is only one niche in the environment. Therefore, fitness is frequency-independent.   

During an avidian's lifespan, it will eventually start to undergo genome replication. As it copies its genome's instructions into a blank daughter genome, some instructions may be copied inaccurately at a point mutation rate set by the experimenter. Additionally, insertion and deletion mutations can occur either during genome replication or during genome division into new daughter genomes. In the experiments performed here, insertion and deletion mutations (indels) were enacted upon genome division. Genome sizes can change every generation by at most 10\% (the default is a maximum change of 100\%). For every indel, two spots in the genome were randomly selected. If the indel was a deletion, everything between those two spots was deleted. If the indel was an insertion, that section of the genome was duplicated. Insertions and deletions occurred at equal frequencies in our experiments.

\begin{figure*}[p] %  figure placement: here, top, bottom, or page
   \centering
   \includegraphics[width=\textwidth]{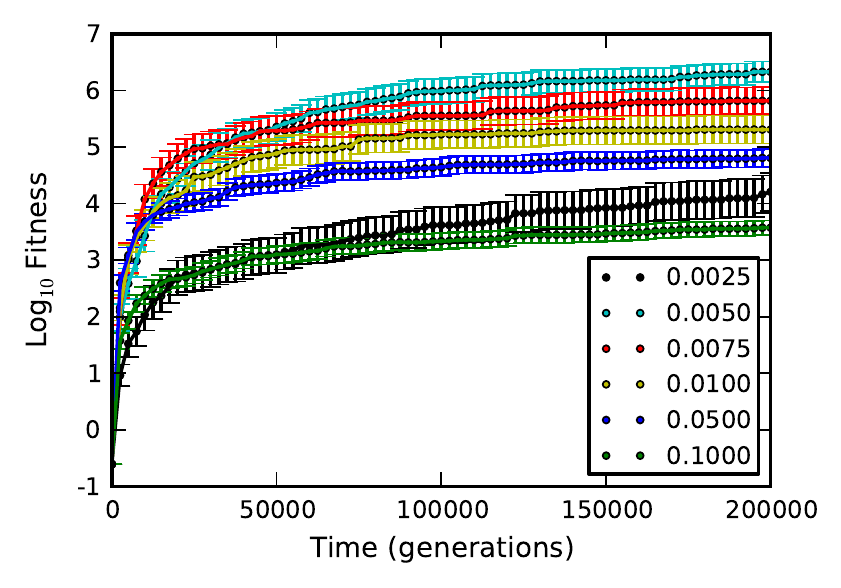} 
   \caption{Increase in population fitness over 200,000 generations is shown for six point mutation rates (0.0025, 0.005, 0.0075, 0.01, 0.05, and 0.1). The $log_{10}$ of fitness is averaged over 100 replicate populations. Error bars represent $\pm$ 1 SE.}
   \label{fig:fitness_allmu}
\end{figure*}

\begin{figure*}[p] %  figure placement: here, top, bottom, or page
   \centering
   \includegraphics[width=\textwidth]{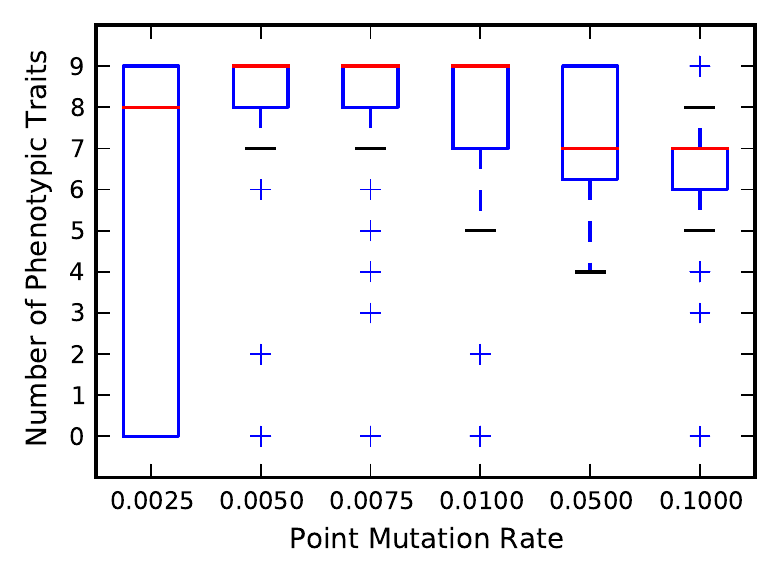} 
   \caption{Number of traits evolved by avidians at six different point mutation rates (maximum number of traits that can be evolved is 9). Red lines are median values from 100 replicate populations, while the upper and lower bounds of the box are the third and first quartile, respectively. Whiskers are either 1.5 times the the quartile value or the extreme value in the data, whichever is closer to the median. Plus signs are outliers.}
   \label{fig:fitness_allmu}
\end{figure*}

\begin{figure*}[p] %  figure placement: here, top, bottom, or page
   \centering
   \includegraphics[width=\textwidth]{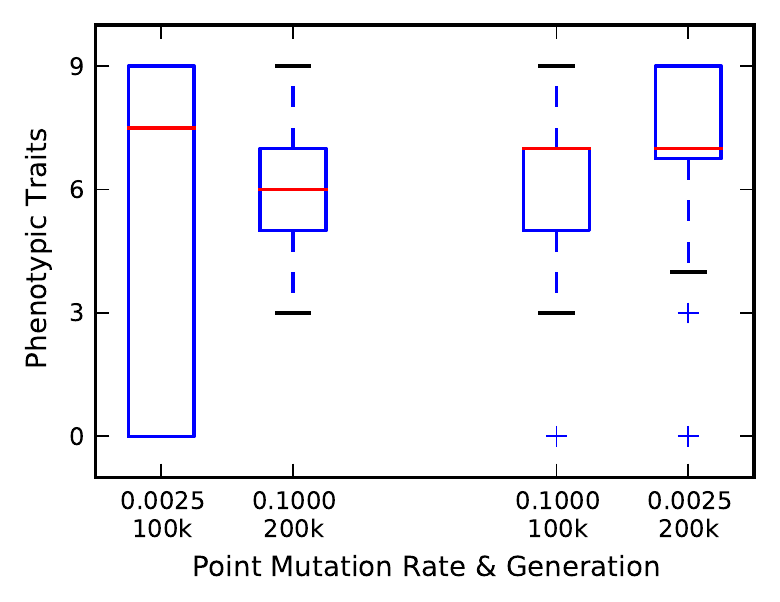} 
   \caption{The number of traits evolved by avidians after mutation rate was switched at the mid-point (at 100,000 generations) in the 200,000 generation long simulations. Left side of the figure shows the statistics for number of traits evolved by the population evolving at point mutation rate of 0.0025, and the number of traits the same population evolved after evolving at mutation rate of 0.1 for 100k generations. Right side shows the reverse scenario: traits evolved by population evolving at point mutation rate of 0.1 at 100k generations, and after mutation rate is switched to 0.0025 for additional 100k generations. Red lines are median values from 20 replicate populations, while the upper and lower bounds of the box are the third and first quartile, respectively. Whiskers are either 1.5 times the the quartile value or the extreme value in the data, whichever is closer to the median. Plus signs are outliers.}
   \label{fig:tasks_switchMu}
\end{figure*}

%\begin{figure*}[p] %  figure placement: here, top, bottom, or page
%   \centering
%   \includegraphics[width=6in]{a_0025_indel_s_hist.pdf} 
%   \caption{Histograms of fitness effects of indels is shown for populations that evolved at point mutation rate of 0.0025. Along the line of descent in 100 replicate populations, there were 19,262 insertions and 16,998 deletions. Insertions (blue bars) are usually beneficial (\textit{i.e.}, fitness effect \textgreater 0), and deletions (red bars) are usually deleterious (fitness effect \textless 0). The two distributions are significantly different (Kolmogorov-Smirnov two-sided test, p-value \textless 1e-100).}
%   \label{fig:indel_s}
%\end{figure*}

%\newpage

\begin{figure*}[p] %  figure placement: here, top, bottom, or page
   \centering
   \includegraphics[width=\textwidth]{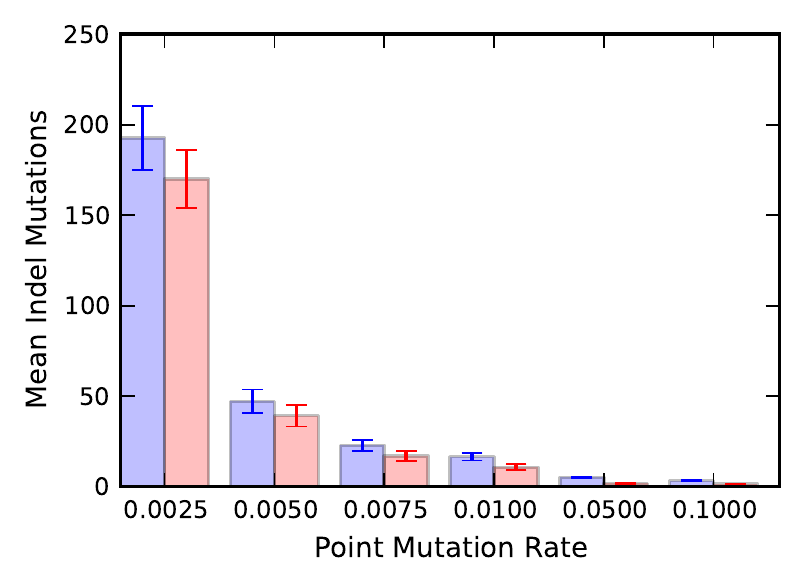} 
   \caption{Insertions (blue bars) are more common than deletions (red bars) at all mutation rates, and frequency of indels decreases as mutation rate increases. Average values over 100 replicates is reported with error bars showing $\pm$ 1 SE.}
   \label{fig:freq_indel_allmu}
\end{figure*}

\begin{figure*}[p] %  figure placement: here, top, bottom, or page
   \centering
   \includegraphics[width=\textwidth]{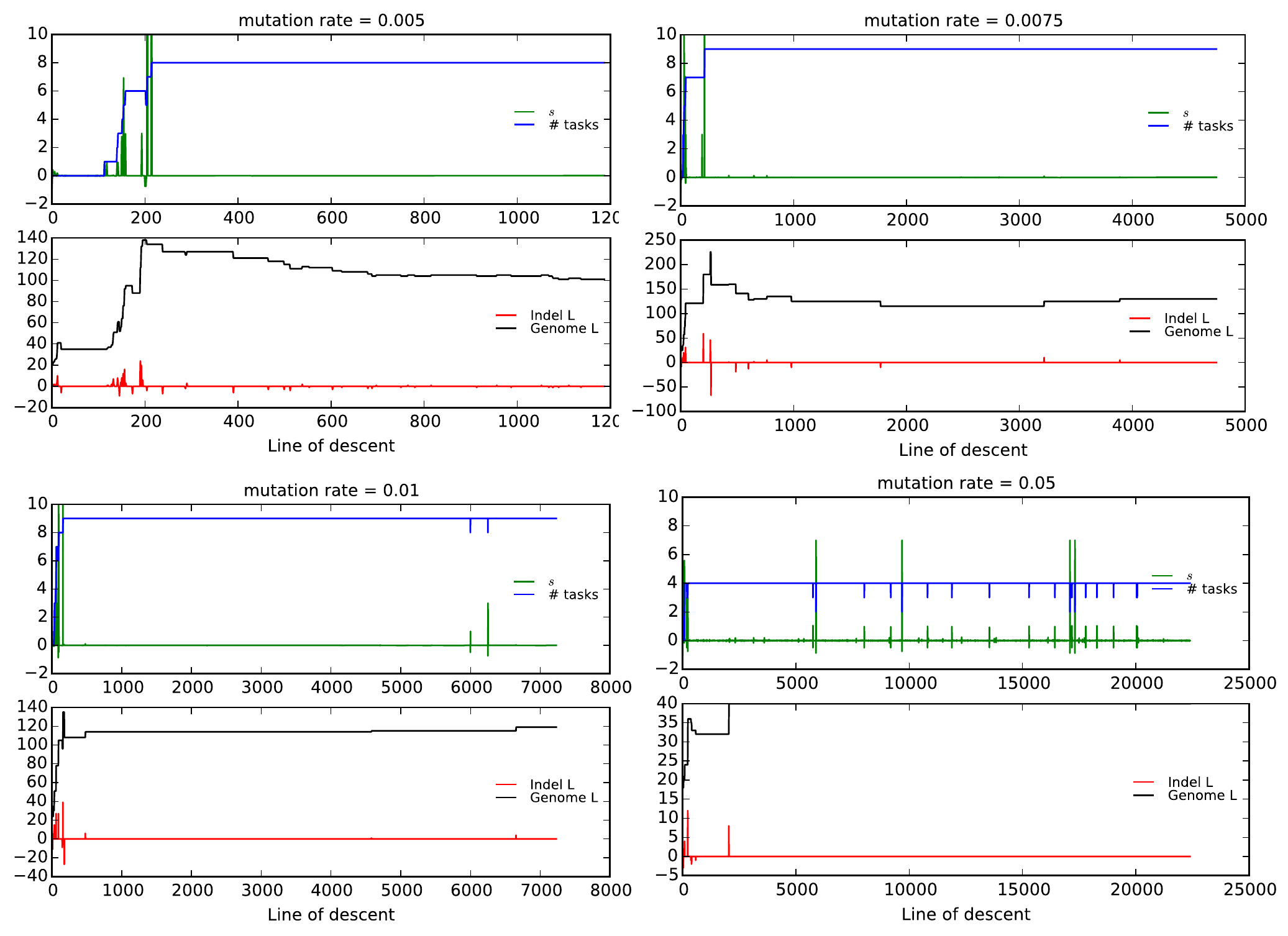} 
   \caption{The line of descent (LOD) of the most fit genome is shown for a single replicate population evolving at the point mutation rates 0.005, 0.0075, 0.01, and 0.05. The fitness effects of genome edit events (insertions, deletions, base substitution) is shown in green, the number of traits evolved over time is shown in blue, the size of indels is shown in red, and the genome length is shown in black.}
   \label{fig:LODmaps}
\end{figure*}

\end{document}